\documentclass[conference]{IEEEtran}

\usepackage{color}

\usepackage{soul}
\usepackage{authblk}
\usepackage{gensymb}
\usepackage[font=footnotesize]{caption}
\usepackage{graphicx}
\usepackage{subfigure}

\ifCLASSINFOpdf
\else
\fi

\usepackage[pscoord]{eso-pic}

\usepackage{amsmath, amssymb, amsthm}
\usepackage{array}
\usepackage{color}
\usepackage{subfigure}
\usepackage{tikz}
\usepackage{textcomp}
\usepackage{hyperref}
\usepackage{lipsum}

\newcommand\copyrighttext{%
  \footnotesize \textcolor{blue}{\textcopyright 2019 IEEE. Personal use of this material is permitted.  Permission from IEEE must be obtained for all other uses, in any current or future media, including reprinting/republishing this material for advertising or promotional purposes, creating new collective works, for resale or redistribution to servers or lists, or reuse of any copyrighted component of this work in other works}}
\newcommand\copyrightnotice{%
\begin{tikzpicture}[remember picture,overlay]
\node[anchor=south,yshift=10pt] at (current page.south) {\fbox{\parbox{\dimexpr\textwidth-\fboxsep-\fboxrule\relax}{\copyrighttext}}};
\end{tikzpicture}%
}

\begin{document}

\title{Using Blockchain to Rein in The New Post-Truth World and Check The Spread of Fake News}

\author{Adnan Qayyum}
\author{Junaid Qadir}
\author{Muhammad Umar Janjua}
\author{Falak Sher}
\affil{Information Technology University (ITU), Lahore, Pakistan}

\maketitle 

\copyrightnotice

\begin{abstract}
\sloppy 
In recent years, ``fake news'' has become a global issue that raises unprecedented challenges for human society and democracy. This problem has arisen due to the emergence of various concomitant phenomena such as (1) the digitization of human life and the ease of disseminating news through social networking applications (such as Facebook and WhatsApp); (2) the availability of ``big data'' that allows customization of news feeds and the creation of polarized so-called ``filter-bubbles''; and (3) the rapid progress made by generative machine learning (ML) and deep learning (DL) algorithms in creating realistic-looking yet fake digital content (such as text, images, and videos). There is a crucial need to combat the rampant rise of fake news and disinformation. In this paper, we propose a high-level overview of a blockchain-based framework for fake news prevention and highlight the various design issues and consideration of such a blockchain-based framework for tackling fake news.
\end{abstract}


\section{Introduction}

The problem of ``\textit{fake news}'' has recently emerged as a big headache for consumers of news. The term ``fake news'', which has now become a buzzword through references to it by high-profile politicians, refers to news that is verifiably false and is intentionally generated to mislead readers  \cite{allcott2017social}. Many people are now referring to the current era as the ``\textit{post-truth}'' era \cite{flintham2018falling}, with experts pointing to two controversially shocking elections in 2016 as the dawn of this era \cite{ball2017post}. Indeed, it appears that we are living in the age of unprecedented misinformation. While fake news has attained its prominence in political discourse, this malaise more generally to any false or misleading information that is purposely spread to deceive someone. 

Fake news has serious consequences for our ideals of democracy, liberty, and society \cite{sunstein2018republic} since fake news can be used to influence people for political purposes and to fulfill agendas that are not necessarily pro-social  \cite{allcott2017social}. We can see that fake news is not only a nuisance but can lead to great harm by noting examples such as the mob lynching in India that resulted from the dissemination of false messages on the social networking site WhatsApp and resulted in the killing of many innocents \cite{gowen2018mob}. There are many similar examples worldwide where the false information was promulgated for deceiving public and society. 

Fake news can be thought of as an instance of yellow journalism in which deliberate hoaxes are created and disseminated across traditional media or social media. Such hoaxes contain sensational, eye-catching, and fabricated content that is intentionally created to mislead, harm, and accomplish certain political and financial objectives. Usually, such content is unwittingly shared by social media users and also by some journalists who want to report news on social media in real time.

While the phenomenon of fake news itself is not new, with propaganda and rumor mongering always being present in human discourse, recent development such as the ease of generating and disseminating digital content, the use of social networking websites, and the development of artificial intelligence (AI) and machine learning (ML), in particular deep learning (DL) based ML methods, for generating fake content has greatly exacerbated this menace. The enormity of the challenge can be seen from the fact that even the big companies like Google and Facebook seem to be stumped by the fake news storm and are far from having solved this problem\footnote{\url{https://tinyurl.com/googleFBfakenews}}. The information verification task, which is already a complicated task in traditional media \cite{mullainathan2005market}, becomes more intransigent when we consider the decentralized peer-to-peer news generation models common on the Internet and the evolved editorial norms and constraints on information dissemination \cite{lazer2018science}.  In the paper \cite{lazer2018science}, the authors have suggested that to avoid the prevalence of fake news, we need to focus more on the original source of the news and emphasize the publishers rather than the individual stories. Another factor that is fueling fake news production and dissemination is the use of powerful AI-based social bots for automating the spread of information and for influencing political perceptions and inclinations \cite{shao2017spread}. 


In addition, news is being increasingly consumed via personalized and customized news sites (e.g., Google's personalized search and Facebook's personalized newsfeed. This leads to the formation of ``\textit{filter bubbles}'' \cite{pariser2011filter}, which refers to a status of intellectual isolation of users from information that disagrees with their viewpoints in which the users are increasingly shown news that is consistent with their bias, beliefs, and worldview, which are guessed through the digital data available about a user's demographics and previous likes/dislikes. This provides a breeding ground and a market for fake news in which false information spread rapidly than the truth as polarized users are fed news that they will likely believe without even cross-checking due to the human cognitive bias known as confirmation bias, due to which users are likely to believe, and less likely to cross check, news that is concordant to their preexisting beliefs and opinions. 




Recent technological advancements in ML and DL, and in particular the development of generative models such as generative adversarial networks (GANs), allow users to imitate reality and create eerily realistic fake audio and video.\footnote{\url{https://tinyurl.com/NPRfakeaudiovideo}} GANs have already been used to create photo-realistic images of fake celebrities, forged signatures, fake audios and vidoes. The use of deep learning based generative models have led to the development of the  ``\textit{deep fake}'' technology that enables one to effectively create the audio and the video of a real person speaking and performing in a way that the person never actually did \cite{chesney2018deep}. This technology can clearly cause much harm to the society including the undermining of journalism and public safety, the manipulation of elections, and the distortion of democratic discourse \cite{chesney2018deep}. Consider the damage one can cause if one can create a deep fake video depicting a person committing a crime, consuming drugs, saying unpopular or even blasphemous things. One could end a politician's career by generating a fake video in which the politician engages in anti-state rhetoric using technology such as the one used in \cite{suwajanakorn2017synthesizing} to create fake videos of US president Obama lip-syncing to arbitrary text. More perversely, one could create fake sex videos of some person creating a horrifying virtual reality ordeal for that person. Deep fakes and related technologies thus portend a looming challenge for privacy, democracy, national security, and ethical norms \cite{chesney2018deep} and there is an urgent need to do research on the technical, legal, and ethical challenges that await the development of this field. 





Many people are looking towards \textit{blockchain} technology, which can enable trust in a peer-to-peer (P2P) network in a decentralized fashion without the presence of a central managing authority, as a potential tool that can help combat the rising challenge of fake news and help build trust in authentic content disseminated on social media. Blockchain technology has recently revolutionalized many businesses by developing solutions that can provide trust, security, and privacy. Blockchain provides a mechanism for maintaining distributed tamper evident ledger to record the transactions in a data structure named blockchain, hence the name blockchain. It is also sometimes referred to as distributed ledger technology (DLT) due to the distributed ledger which is maintained using the distributed consensus algorithm by networks peers. Blockchain has emerged as one of the technological breakthroughs in recent years and has disrupted the banking system and it has been used beyond its foremost application in cryptocurrencies, e.g., real estates and fraud detection in online business. Blockchain enabled cryptocurrencies, e.g., Bitcoin \cite{nakamoto2008bitcoin} and Ethereum enforce security of the transactions in a decentralized fashion that prevents tampering of the systems state without involving any central bank or financial institution. 

The \textit{major contribution of this paper is} that we provide a blockchain framework to counter the fake news problem. We also provide a comprehensive overview of the fake news phenomenon and describe its prevalence and the challenges it poses. To address the fake news challenge, we propose a smart contract-based blockchain framework and describe its overarching structure in this paper. 

\section{Background}
\label{sec:back}

\subsection{Fake News Related Background}

\subsubsection{The Science of Fake News}

To deeply understand fake news, a recent paper \cite{lazer2018science} has attempted to develop the science of fake news and thereby answer questions such as (1) why and how fake news are created? and (2) what are the essential elements that make such content spread virally? The authors argued that there are few scientific answers to the basic questions about the impact and prevalence of the fake news and the existing solutions are not commensurate to the enormity of this problem. Further, they emphasized that the rising epidemic of fake news can only be combated with an interdisciplinary effort and there is a need to redesign the information ecosystem that can values and promotes truth in current digital era. 
Given the power of social media to make any information a viral one, detection and prevention of fake news and false information is required and crucial. However, the rapid increase in fake news highlights that the long-standing organizations are lagging behind in combating this global problem. 





\subsubsection{Solutions Proposed For Curbing Fake News}

Traditional methods for detecting fake news are not applicable and effective for online social networks \cite{shu2017fake}. Therefore, various approaches have been presented in the literature using modern tools and techniques to tackle the problem of fake news on social media platforms.  
A network analysis approach for detection and mitigation of fake news is presented in \cite{shu2019studying}. In a similar study \cite{jin2016news}, authors build credibility network with conflicting relations for detection of fake news spread via tweets. 
In \cite{tacchini2017some}, an approach for fake news detection is presented using machine learning techniques and Boolean crowd-sourcing algorithms. 
A project named ``Solid''\footnote{\url{https://solid.mit.edu/}} has been initiated by the inventor of world wide web (www)--Prof. Tim Berners-Lee that aims to develop decentralized social applications while introducing privacy preserving authenticity of data ownership. In addition to the work of researchers and academicians, there are lots of efforts being done by big companies to mitigate the fake news on social networks. Recently, Google announced \textit{``Google News Initiative''} for putting their efforts to support news industry in combating rampant of fake news\footnote{https://tinyurl.com/GoogleResponseFakeNews}. Similarly, Facebook\footnote{www.facebook.com/zuck/posts/10103269806149061?pnref=
story} and Wikipedia\footnote{https://tinyurl.com/GuardianWikiTribune} are also working towards developing a feasible solution for mitigating outbreak of disinformation.   

A line of research focused on using blockchain for fake news identification and prevention is also catching up and there are very few studies which mostly provide the theoretical approaches to tackle this problem. We discuss these at the end of the next subsection.





  





\subsection{Blockchain Related Background}



\subsubsection{Cryptographic Hashing}

A hash function is a unidirectional cryptographic function that accepts an input of any size and returns a fixed size output. A cryptographic hash function should encompass three essential security properties namely, hiding, collision-free, and puzzle friendliness. Hiding property implies that if the output of a hash function is known, it is infeasible to find its corresponding input value and collision-free property ensures that for two different input values the hash output should be different. While puzzle friendliness property is used to create a mathematical search puzzle which aims at searching for a solution in a very large space. 


\subsubsection{Blockchain and Hash Pointers}

A blockchain is a linked list built using hash pointers where a hash pointer is simply a pointer that gives the location of the stored information including its cryptographic hash. In blockchain, a pointer to the previous block is a hash pointer that contains a digest of that block. 
Thus the blockchain data structure provides a tamper-evident log of the stored transactions. For example, after modifying the transactions in a block, an adversary has to change all hash pointers in all successive blocks, which is practically impossible in blockchain that is linked together using hash pointers. 


\subsubsection{Digital Signatures}

Digital signatures are primitive used in cryptography along with hash functions. A digital signature provides a digital analogous to the handwritten signatures and has two essential properties, i.e., only you can make your signatures and no one can copy them (i.e., unforgeable). In a blockchain based system such as Bitcoin, digital signatures are made using secret and public key pair, where a sender uses the secret key to sign a transaction and receiver and any participating node in the P2P network can verify that it is signed under the sender's public key. 


\begin{figure*}[h!]
\centering
\includegraphics[scale=0.5]{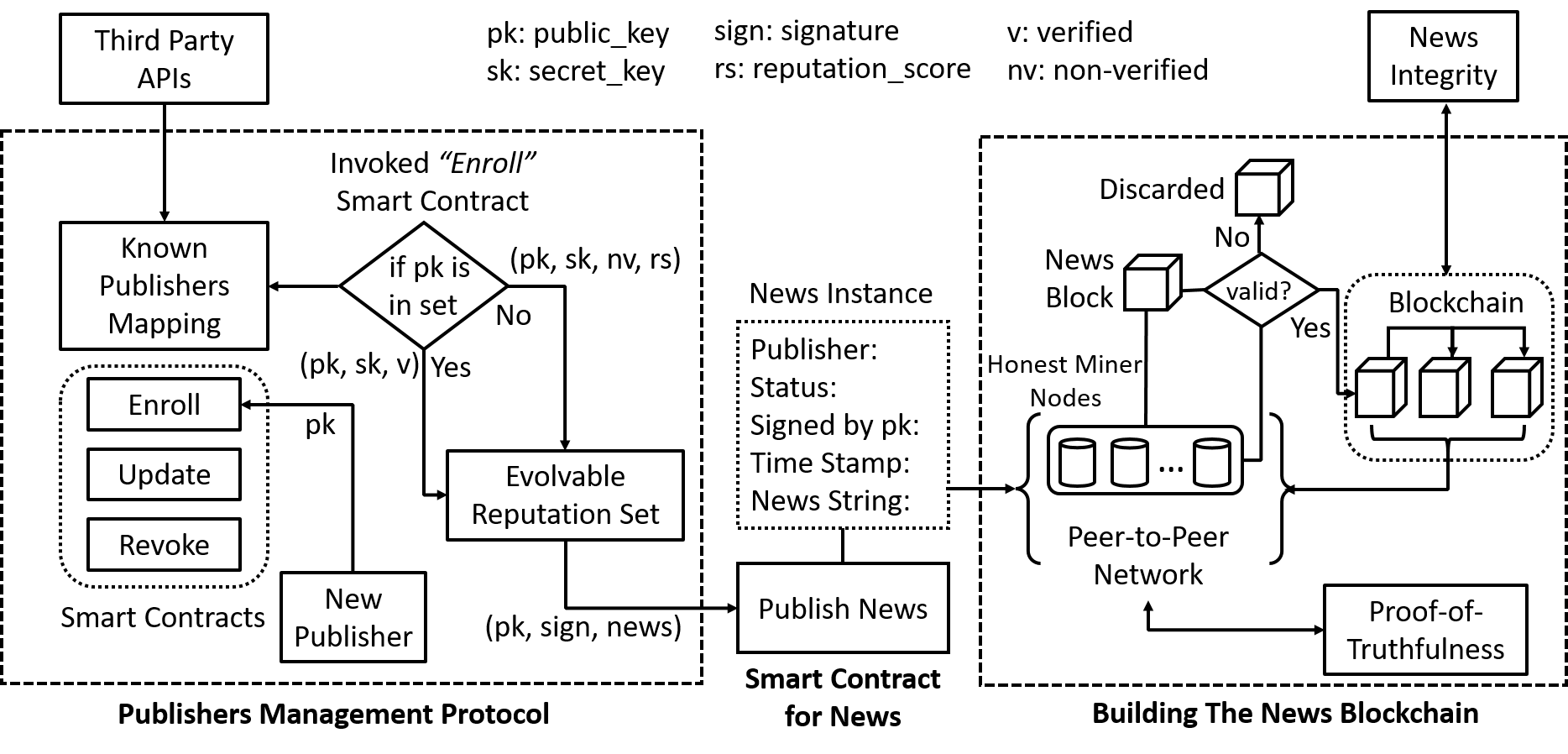}
\caption{The proposed architecture of the blockchain-based news verification framework has three components: (1) Publisher Management Protocol; (2) Smart Contract for News; and (3) The News Blockchain.}
\label{fig:sys}
\end{figure*}


\subsubsection{Distributed Consensus}

The backbone of blockchain-based cryptocurrencies is distributed consensus. A distributed consensus protocol should work in a situation where some of the participating nodes are malicious or faulty and should have the following two properties: 1) all honest nodes should terminate on the same value, and 2) that value must have been proposed by an honest node. The distributed consensus protocol used by Bitcoin is proof-of-work (PoW) which is described as follows. 


\textit{Proof-of-Work (PoW)}: The PoW consensus protocol was familiarized by Bitcoin and the key goal of using PoW is to prevent double spending attack which is accomplished using hash puzzles. Every time a random node is selected to propose a new block and to successfully publish a block, it is required to find a \textit{nonce}. The nonce is a value when concatenated with the previous hash pointer and list of transactions and then hashed together, it should be less than the predefined target as depicted by Eq. \ref{eq:1}. Finding such a value is computationally very intensive and known as mining and the node that successfully find this value is called a miner. A miner node broadcast its new block and other nodes on network accept it if and only if all transactions in that block are valid, unspent, and are signed by valid signatures. The nodes express the acceptance of the block by including its hash in the next block they create.


\begin{equation}
    H(nonce \Vert prev\_hash \Vert tx_1 \Vert tx_2 \Vert ... \Vert tx_n) < target
    \label{eq:1}
\end{equation}


\subsubsection{Blockchain To Check Fake News}

In a recent study \cite{huckle2017fake}, an Ethereum blockchain enabled prototype framework is proposed for verification of content's credibility. The proposed solution uses preservation metadata implementation strategies (PREMIS) for cryptographically storing digital media on the blockchain. In \cite{shang2018tracing}, the authors utilize the idea that the source of news can be traced by keeping a record of time stamp service and the chain connections between the blocks and consequently propose a blockchain-based approach for decentralized distributed storage for tracing the origin of the news. In another related study \cite{jing2018theoretical}, the authors presented a theoretical framework that uses blockchain to prevent fake news dissemination. The key concern of the authors is to devise a hybrid algorithm using the concepts and principles of blockchain for increasing transparency and credibility of the content spreading on social networks. 

  





\section{Fake News Prevention using Blockchain: A Solution Blueprint}
\label{sec:method}

Blockchain can be leveraged to preserve and verify the integrity of the news and other multimedia content being shared online. In this section, we present the blueprint of a blockchain-based
framework that relies on smart contracts for fake news detection and prevention. We first discuss the various design challenges of developing such a system and then discuss our proposed system.




There are numerous design challenges in developing an effective blockchain-based solution for combating fake news and misinformation, which remain unaddressed in the literature. Various current proposals aiming at using blockchain for fake news or false information prevention (e.g., \cite{huckle2017fake, jing2018theoretical,shang2018tracing}) mainly used hashing but a simple hash based approach may not work in a practical setting, as hashing is very sensitive to noise and can result in a different hash if there is change of a character or even a single bit. This pitfall is acknowledged in \cite{huckle2017fake} as a limitation of their proposed system. 


In this study, we attempt to incorporate these challenges in designing a blockchain based decentralized system for fake news prevention and mitigation. The proposed system architecture is presented in Figure \ref{fig:sys} and is described next.



\subsection{Publishers Management Protocol}
The key function of publishers management protocol is to intelligently distinguish in a decentralized fashion between credible and non-credible sources of news/information. The system uses three types of smart contracts that are used to \textit{enroll}, \textit{update}, and \textit{revoke} the identities of news organizations. Also, we use \textit{status} and \textit{reputation score} for the news entities working on our system (described next).  

\textit{Enrollment Smart Contract}: 
The system has an existing mapping of public keys that are in use by various news media organizations, which can be used to verify their identities in real life. If there is no such key for a specific news entity the system can search the web through third-party APIs. Each time a news entity wants to register, the system verifies its identity by asking it to sign a message with its existing public key. 
If the verification process is successful, the news entity will be given a verified status; otherwise, the news entity can publish but only as a \textit{non-verified} news publisher. In each case, the system assigns a pair of secret and public key to the news entity being enrolled that will be used for digital signatures scheme. 

\textit{Update Identity Smart Contract}:
Any registered news publisher can update its identity and is allowed to obtain multiple identities---e.g., a news channel may want to publish news about sports with a different handle. To get another identity in the form of public and private key pair, a registered news publisher is required to verify its previous identity (i.e., its public key previously registered on the system). The update identity smart contract is used to facilitate such requests.  

\textit{Revoke Identity Smart Contract}: 
The revocation of a smart contract deals with the termination of existing news publishers either on their own request or if the system has identified a certain news publisher to behave anomalously over a specified period of time. This is accomplished by computing a reputation score for quantifying the credibility of a publisher. 

\textit{Evolvable Reputation Set}:
Our system maintains a \textit{reputation set}, which contains a set of authentic and credible news sources. To make this set evolvable, we assign an initial reputation score of zero to each non-verified media outlet, and allow this score to evolve with the passage of time if that news entity shares true news or the news posted by the verified media outlets. If a non-verified news outlet manages to get a specified reputation score in a given time period, it will get a status of the verified source; otherwise, if a non-verified news outlet spreads fake news and false information, its identity will be revoked after that time period from the system. Another way to maintain an evolvable set is to get feedback from the consumers of the news, but this introduces the problem of subjectivity and bias and also opens the system to the risk of malicious actors.

\subsection{Smart Contract for News}

\subsubsection{News Publishing}
The \textit{create news smart contract} is used to publish the news on the network. It can be invoked by the account(s) of media outlets willing to publish any news by giving their public key and digitally signed news. The smart contract will store the relevant information such as publisher name, status, public key, time stamp, and news string in a structure and will broadcast the news instance to the P2P network.  

\subsubsection{News Integrity} 
Ensuring the integrity of the news posted by two different media entities about a specific topic is another challenge. It has already been discussed that a simple hash based approach is not suitable for ensuring news credibility due to its lack of robustness. To overcome this issue, we propose the use of \textit{semantic similarity} of a news posted by two or more different news outlets. The semantic similarity can be used by the system to gauge the integrity of a news article by checking for news on the blockchain (i.e., whether it has been posted by a verified news organization or not). This semantic similarity index can be measured by a tool such as word embeddings (e.g., word2vec) or other advanced ML methods through which contextual similarity can be computed between words and across documents. 

\subsection{Building The News Blockchain}

\subsubsection{Role of Honest Miner Nodes}

Since there can be malicious nodes in the P2P network that may want to modify the content for spreading false information, the proof-of-authority (PoA) consensus protocol can be used by miner nodes for maintaining the blockchain while proposing a new ``news block''. With the assumption that a majority of honest miner nodes are providing their services to maintain the system's integrity, it is more likely that each time an honest node gets a turn to propose a new news block that this block should be added to the blockchain. The honest miner nodes can be deployed by credible mainstream media outlets themselves or any other trusted entity having an aim to ensure the integrity of the news being shared on the network.  

\subsubsection{Proof-of-Truthfulness (PoT)}
Any participating node in the network can verify whether the news it has encountered is part of the blockchain or not by using the proof-of-truthfulness (PoT) method. This is done by storing the news in a Merkle tree,  which is a binary tree build using hash pointers in which nodes at $n-1$ level contains the hash pointers to the news stored at $n^{th}$ level and the root (considered level 0) contains a hash pointer to the two nodes at level 1. Given a news, anyone can verify its truthfulness in $O(\log(n))$ time by searching only a single branch of the tree from that news to the root. 

\subsection{Framework Illustration}
In this section, we provide an illustration of the proposed framework. When a new news publisher requests to join the system, the \textit{enrolment smart contract} will be invoked and will check whether this publisher's public key is in existing mapping or not. Based on the outcome, the publisher will be assigned a public and secret key pair along with a status of either verified or non-verified publisher and an initial reputation score then it will become part of evolvable reputation set. During the period an existing publisher can update or revoke its identity and the identity of a misbehaving publisher will be revoked automatically as described above. Whenever a publisher wants to publish, \textit{create news smart contract} will be invoked to facilitate the news publishing and it will put the news along with other required parameters in a block and will broadcast it to the P2P network. In the P2P network, the miner nodes will put this block on the blockchain if it is identified as a valid block. Once the news block becomes the part of the blockchain, news integrity and truthfulness can be verified using semantic similarity and Merkle tree respectively as described in earlier sections.




\section{Discussion}
\label{sec:dis}


There are numerous open issues in developing a solution to the problem of fake news and fake content. 


\begin{itemize}
  
  \item An important future work is to pursue the development of a decentralized distributed trust framework using blockchain that ensures the authenticity and credibility of news sources of heterogeneous quality without requiring a central authentication/trust infrastructure. Existing blockchain archetypes developed for various cryptocurrencies cannot be directly adopted for the development of a system aiming at fake news prevention due to the unique challenges associated with news verification.

  \item  It is important to incorporate the contextual knowledge for verifying the news integrity. An interesting line of future work will be the use of AI, in particular natural language processing (NLP) methods, for developing deep insights about the similarity of news and using this information to quantify the trustworthiness of news items. 

  \item Preserving the privacy and security of individuals' personal photos that they shared on their social media account also demands urgent attention since there is a possiblity that some one may use these personal photos to train an ML model such as ``deep fakes'' to create fake content. A possible blockchain-based solution can be to cryptographically store the images and videos on blockchain in such a way that every interaction with the content is detectable and tractable.  
  




\end{itemize}


\section{Conclusions}
\label{sec:con}
The progress made by artificial intelligence (AI) techniques for customization, dissemination, and generation of content and the increasingly digital habitat of human lives has created the unhappy situation where sensationalized fake content and misinformation thrives and spreads like wildfire; whereas establishing the authenticity of the truth and differentiating reality from fakes is becoming increasingly daunting. This creates a number of imposing technical, legal, and ethical challenges. The technology of blockchain, being a decentralized ledger technology, promises to bring transparency and trust to this new ``post-truth'' world by enabling features such as smart contracts, decentralized consensus, and tamper-proof authentication. In this paper, we have introduced the modern malaise of realistic-looking fake content and focused specifically on the phenomena of fake news. To address the fake news problem, we have proposed a blockchain-based framework for detection and mitigation of fake news and have described a high-level blueprint of our solution. In our future work, we aim to extend this work to develop an actual prototype of such a system. 

\bibliographystyle{plain}

\end{document}